\documentclass[reprint,amsmath,amssymb,aps,superscriptaddress,longbibliography]{revtex4-2}

\usepackage{physics}
\usepackage[T1]{fontenc}
\usepackage[utf8]{inputenc}
\usepackage{multirow}
\usepackage{amsmath}
\usepackage{mathtools}
\usepackage{graphicx}
\usepackage{bm}
\usepackage{braket}
\usepackage{soul} 
\usepackage{tcolorbox}
\usepackage{url}
\usepackage[linktocpage, colorlinks=true, linkcolor=darkblue, citecolor=darkblue, urlcolor=darkblue]{hyperref}
\definecolor{darkblue}{rgb}{0,0,.5}

\usepackage[normalem]{ulem}
\usepackage{accents}

\begin{document}

\title[]{Benchmarking the Exponential Ansatz for the Holstein model}

\author{Junjie Yang}
\affiliation{Division of Chemistry and Chemical Engineering, California Institute of Technology, Pasadena, California 91125, USA}
\author{Zhi-Hao Cui}
\affiliation{Department of Chemistry, Columbia University, New York, NY 10025}
\author{Ankit Mahajan}
\affiliation{Department of Chemistry, Columbia University, New York, NY 10025}
\author{Huanchen Zhai}
\affiliation{Division of Chemistry and Chemical Engineering, California Institute of Technology, Pasadena, California 91125, USA}
\author{David R. Reichman}
\affiliation{Department of Chemistry, Columbia University, New York, NY 10025}
\author{Garnet Kin-Lic Chan}
\email{gkc1000@gmail.com}
\affiliation{Division of Chemistry and Chemical Engineering, California Institute of Technology, Pasadena, California 91125, USA}

\date{\today}

\begin{abstract}
Polarons are quasiparticles formed as a result of lattice distortions induced by charge carriers. The single-electron Holstein model captures the fundamentals of single polaron physics. We examine the power of the exponential ansatz
for the polaron ground-state wavefunction in its coupled cluster, canonical transformation, and (canonically transformed) perturbative variants across the parameter space of the Holstein model. Our benchmark serves to guide  future developments of  polaron wavefunctions beyond the single-electron Holstein model.
\end{abstract}
\maketitle

\section{Introduction}
\label{sec:Introduction}
Electron-phonon interactions play an essential role in many solid-state materials \cite{wellein1998}.
Polarons are quasiparticles formed due to a strong electron-phonon interaction, which results in the trapping of electrons by localized lattice distortions \cite{landau1933}.  
Polaron phenomenology has been extensively studied through lattice models and semi-empirical Hamiltonians. Out of these, the simplest is probably the Holstein model~\cite{holstein1959}, whose phase diagram captures both the phenomenon of polaron formation and the crossover between small and large polaron regimes.
Despite the model's simplicity, analytical techniques are only applicable under restricted conditions \cite{gerlach1991}. 
Thus most results have been obtained by applying numerical methods, including exact diagonalization (ED) \cite{zhang1998, dagotto1994}, density matrix renormalization group  (DMRG) \cite{jeckelmann1998}, variational wavefunctions~\cite{lamagna1996,brown1997,romero1999,romero1999a,wang2020a},  perturbation theories \cite{gogolin1982,lang1963,stephan1996}, and canonical transformation formalisms \cite{sio2019,lee2021,luo2022}. 


The current work studies and benchmarks the exponential electron-phonon ansatz for the single-electron Holstein model. We consider two specific kinds of exponential ansatz. The first, coupled cluster theory, is widely recognized as a highly accurate method in computational chemistry~\cite{coester1958,cizek1966,paldus1975,cizek1980,bartlett2007,shavitt2009}. Here we use the electron-phonon coupled cluster formalism introduced by some of us in Ref.~\cite{gao2020}. The second is the well-known variational Lang-Firsov ansatz and its recently introduced perturbation theory, as discussed  in Ref.~\cite{cui2023} also by some authors of this work. The variational Lang-Firsov ansatz \cite{lang1963}
can be viewed as a unitary coupled cluster theory, and thus fits within the family of exponential ansatz wavefunctions. We study the numerical performance of these methods across different parameters of the model, benchmarking against converged exact results from density matrix renormalization group simulations
\cite{jeckelmann1998,zhai2021,zhai2023}.
In additional to the energy, we examine multiple observables that characterize polaron formation and the transition between large and small polarons.

\section{Theory}
\label{sec::Theory}
\subsection{Holstein Model}
\label{subsec::Holstein-Model}
The single-electron Holstein model defines a minimal spinless one-electron lattice model containing interactions between a single electron band and a set of local phonons.   
The Hamiltonian for a $L$-site one-dimensional Holstein model is
\begin{equation}
\begin{aligned}
    H = & - t \sum_{\langle lk \rangle}  \left(a_{l}^{\dagger} a_{k} + a_{k}^{\dagger} a_{l}\right) 
    + \omega \sum_l b_l^{\dagger} b_l \\
    & + g \sum_l \rho_l \left(b_l^{\dagger} + b_l\right)
\end{aligned}
\end{equation}
where $a_{l}^{\dagger}$ ($a_{l}$) creates (annihilates) an electron on site $l$, and $b_l^{\dagger}$ ($b_l$) creates (annihilates) a phonon of 
frequency $\omega$ on site $l$, $\rho_l=a_{l}^{\dagger} a_{l}$ is the electron number operator on site $l$, and periodic boundary conditions are assumed. 
The electron-phonon
coupling parameter is denoted as $g$, and
we set the hopping parameter $t$ to be $1$ (thus all energy values will implicitly be in units of $t$).

The physics of the Holstein model can be understood in different limiting cases. Defining the adiabaticity parameter as $\omega$,  $\omega \ll t$ corresponds to a slow response of the lattice distortion to electron hopping.
In this scenario, the Born-Oppenheimer approximation is valid and the electronic motion is consequently modified by quasi-static lattice deformations (i.e. an adiabatic potential surface).
Alternatively, when $\omega \gg t$ (the anti-adiabatic limit), the lattice deformation adapts instantaneously to the electron's position.
Then, only the vibrational ground state is  involved in low-energy 
electron hopping processes. 

Another axis along which to understand the physics is with respect to the electron-phonon coupling $g$ itself.
In the weak coupling regime, defined by $g/\omega \ll 1$ and $g^2 / \omega \ll 2t$, the system resembles a quasi-free electron dragging a phonon cloud: {this is known as a large polaron}. 
In contrast, in the strong coupling regime, the electronic position is closely correlated with the lattice distortion it generates: {this is known as a small polaron, which is referred to as self-trapped}. Although some early analytical perturbation theory  
suggested a sharp transition between the two~\cite{alexandrov1994,marsiglio1995}, it has been established from rigorous analysis that there is a smooth crossover between the small and large polaron limits~\cite{gerlach1991}.
The accurate description of the self-trapping crossover represents a challenge for most computational methods~\cite{stephan1996}. 

The exact ground-state of the Holstein model can be expressed as
\begin{equation}
\label{eq::fci}
    |\Psi_0\rangle = \sum_{I} C_{I} |\Phi_I\rangle
\end{equation}
where $|\Phi_I\rangle$ is a configuration of the electron-phonon system,
\begin{equation}
\label{eq::fci-config}
    |\Phi_I\rangle = |\sigma_1^I n_1^I \rangle\otimes |\sigma_2^I n_2^I \rangle \otimes \cdots \otimes |\sigma_L^I n_L^I \rangle
\end{equation}
with $\sigma_l^I$ an electronic state on site $l$, which can be either empty or singly occupied, 
and $n_l^I$ is the phonon occupation state on site $l$, which is a number between $0$ and $N$ where $N$ is the maximum local phonon number. 
The number of configurations in the above is $L \times (N+1)^L$, and $N$ should be taken to infinity for converged results. The rapid growth in the number of configurations with both $N$ and $L$ limits the usefulness of exact diagonalization (ED) and motivates the study of approximate ansatzes. We focus on two variants of the exponential ansatz for the ground-state: the coupled cluster method and the Lang-Firsov transformation. 
For benchmarking purposes, we employ 
the density matrix renormalization group (DMRG), which provides highly accurate results for finite $N$, but where $N$ must be extrapolated: this is described in  Appendix~\ref{appendix::eph-dmrg}.


\subsection{Reference states}

The coupled cluster and Lang-Firsov methods require a starting reference state. We will assume the reference state to be a product state of a single-particle electronic orbital and a site-dependent coherent state, 
\begin{equation}
\label{eq:mfansatz}
\begin{gathered}    
    |\Phi_0\rangle = |\Phi_0^{\mathrm{F}}\rangle
    \otimes |\Phi_0^{\mathrm{B}}\rangle \\
    |\Phi_0^{\mathrm{F}}\rangle = \sum_{l} A_{l} a_{l}^{\dagger} |0^{\mathrm{F}}\rangle  \\
    |\Phi_0^{\mathrm{B}}\rangle = 
    \mathrm{exp}\bigg[\sum_l {\xi_l b_l^\dagger 
    - \xi_l^* b_l}\bigg] | 0^{\mathrm{B}} \rangle
\end{gathered}    
\end{equation}
where $|0^\mathrm{F}\rangle$ and $|0^\mathrm{B}\rangle$ are the physical vacua for electrons and phonons respectively.

In the simplest case, which we refer to as the mean-field (MF) reference state, $A_l$ and $\xi_l$ are chosen to minimize the energy $\langle \Phi_0 | H |\Phi_0\rangle / \langle \Phi_0 | \Phi_0\rangle$, with 
$\xi_l = {g \langle \rho_l \rangle} / {\omega}$ and $\langle \rho_l\rangle = 
A_{l}^* A_{l}$. $\xi_l$ represents the equilibrium  shift of the phonon mode at site $l$ induced by the electron density. The optimal $A_l$ either satisfies translational invariance (delocalized) or it breaks translational invariance (localized). These two solutions, $|\Phi_0^{(\mathrm{D})}\rangle$, $|\Phi_0^{(\mathrm{L})}\rangle$ respectively, are favoured in different regimes of the Holstein model parameter space, and can be viewed as a mean-field description of the self-trapping crossover. 
However, it is known that the mean-field description is not accurate in this region, and
an important test of the more sophisticated exponential ansatzes we explore on top of these mean-field reference states is to see how well they improve the description of the crossover.

An alternative is to reoptimize and improve $\mathbf{A}$ and $\boldsymbol{\xi}$ in the presence of the exponential ansatz parameters. We discuss this option below.

\subsection{Coupled Cluster Models for Electron-phonon Systems}
\label{subsec::eph-cc}

\begin{table*}[ht!]
    \centering
    \renewcommand{\arraystretch}{0.0} 
    \setlength{\tabcolsep}{2.00em}
    \begin{tabular}{ccccc}
    \hline\hline
        model &  $T_{\mathrm{e}}$ & $T_{\mathrm{p}}$ & $T_{\mathrm{ep}}$ & scaling \\ \hline
        & & & & \\
        CCS-1-S1 & $t_{ai} a^\dagger_a a_i$ & $ t_l b_{l}^{\dagger}$ & 
        $t_{l,ai} b_l^{\dagger} a_a^{\dagger} a_i$ & $L^3$ \\
        & & & & \\
        CCS-2-S2 & $t_{ai} a^\dagger_a a_i$ & $ t_l b_{l}^{\dagger} + 
        \frac{1}{2!} t_{lk} b_{l}^{\dagger} b_{k}^{\dagger}$ & 
        $t_{l,ai} b_l^{\dagger} a_a^{\dagger} a_i + \frac{1}{2!} t_{lk,ai} b_l^{\dagger} b_k^{\dagger} a_a^{\dagger} a_i$ 
        & $L^4$ \\
        & & & & \\
        LF-HF & 0 & 0 & $\lambda_l a_l^{\dagger} a_l (b_l^{\dagger}  - b_l)$ & $L^3$ \\
        & & & & \\
    \hline\hline
    \end{tabular} %
    \caption{The exponential ansatzes employed in this study; the wavefunctions are of the form $e^{T_{\mathrm{ep}}} e^{T_\mathrm{e}} e^{T_\mathrm{p}} | \Phi_0\rangle$, where $|\Phi_0\rangle$ is the mean-field electron-phonon reference state.
    Computational scaling is described as a function of the number of sites $L$. 
    Summation symbols are omitted for brevity.
    }
    \label{tab::models}
\end{table*}

The coupled cluster method for electron-phonon systems is formulated based on the
exponential wavefunction {ansatz} \cite{white2020},
\begin{equation}
|\Psi_{\mathrm{CC}}\rangle = \exp(T_{\mathrm{CC}}) \; |\Phi_0\rangle
\end{equation}
where the excitation operator $T_{\mathrm{CC}}$ consists of an electronic part, a 
 phononic part, and a coupling:
\begin{equation}
\label{eq::t-and-tau}
\begin{aligned}
T_{\mathrm{CC}} & = T_{\mathrm{e}}+T_{\mathrm{p}}+T_{\mathrm{ep}} = \sum_{\mu} t_{\mu} \tau_{\mu}\\
\tau_\mu & \in \lbrace a^\dagger_a a_i, b^\dagger_l, b^\dagger_l a^\dagger_a a_i \cdots 
\rbrace
\end{aligned}
\end{equation}
where $i$, $a$ are occupied and virtual electronic orbital indices, and $l, k, m, n$ are used for site indices for the boson operators.
The truncation of the excitation operator determines the accuracy and cost scaling of the method. Given the one-electron nature of the Holstein model, there is only a single occupied electronic orbital (i.e. a single index value for $i$, with the form of the orbital specified by $\mathbf{A}$ in Eq~\ref{eq:mfansatz}), and the electronic excitations are captured exactly using singles excitations of the form $a^\dag_a a_i$, thus we truncate the electronic excitation to this level, referred to as singles.
The order of the phonon and coupling excitations is then denoted by X-SY, where X, Y are numbers that show the excitation order. For example, CCS-1-S1 indicates single electronic excitations, single phonon excitations, and single coupled excitations, corresponding to the excitation operator $\sum_{ia} t_{ai} a^\dag_a a_i + \sum_l t_l b^\dag_l + \sum_{l,ai} t_{l,ai} b_l^{\dagger} a_a^{\dagger} a_i$. 
The formulae and the scaling of the CC models considered in this work are shown in Table~\ref{tab::models}.
(It is helpful to recognize that the models we consider are invariant to shifts of the boson operators e.g. $b_l \to b_l + \xi_l$ and  $b_l^\dagger \to b_l^\dagger + \xi_l^*$, in the sense that any such shift can be absorbed into a redefinition of the coupled cluster amplitudes).

The amplitudes are obtained by solving the projected Schr\"{o}dinger equation:
\begin{equation}
\label{eq::cc-equations}
    0 = \langle \Phi_\mu | \mathrm{e}^{-T_{\mathrm{CC}}} H \mathrm{e}^{T_{\mathrm{CC}}}
    |\Phi_0 \rangle \quad |\Phi_\mu\rangle = \tau_\mu |\Phi_0\rangle
\end{equation}
and the energy is obtained from
\begin{equation}
    E_{\mathrm{CC}} = \langle \Phi_0 | \mathrm{e}^{-T_{\mathrm{CC}}} H \mathrm{e}^{T_{\mathrm{CC}}} | \Phi_0 \rangle
 = \langle \Phi_0 | {H}_{\mathrm{CC}}(\mathbf{t}) |\Phi_0\rangle
\end{equation}
where ${H}_{\mathrm{CC}}(\mathbf{t}) = \mathrm{e}^{-T_{\mathrm{CC}}} H \mathrm{e}^{T_{\mathrm{CC}}}$ is the coupled cluster effective Hamiltonian.
The same equations have also been used in coupled cluster theories for cavity polaritons that have been independently developed 
in Refs.~\cite{mordovina2020,haugland2020,li2023}.
In this study, all the coupled cluster equations were formulated using the \textsc{Wick} package and solved using the Newton-Krylov method,
which approximates the inverse Jacobian matrix using the Krylov subspace method \cite{yang2020,knoll2004}. Note that as all matrix elements are evaluated using Wick's theorem, there is no need to truncate the phonon number.

Because the energy is defined from an asymmetric expectation value, the coupled cluster energy is not necessarily variational. In addition, it does not satisfy a Hellman-Feynman theorem, thus 
in order to obtain observables other than the energy, we instead define a coupled cluster Lagrangian
\cite{shavitt2009},
\begin{equation}
\begin{aligned}
    \mathcal{L}_{\mathrm{CC}}(t_\mu, \lambda_\mu) & = E_{\mathrm{CC}}
    + \sum_{\mu} \lambda_\mu \langle \Phi_\mu |\mathrm{e}^{-T_{\mathrm{CC}}} H \mathrm{e}^{T_{\mathrm{CC}}} | \Phi_0 \rangle \\
    & = \langle \Phi_0 | (1 + \Lambda_{\mathrm{CC}}) \; \mathrm{e}^{-T_{\mathrm{CC}}} H \mathrm{e}^{T_{\mathrm{CC}}} | \Phi_0 \rangle
\end{aligned}
\end{equation}
where $\lambda_\mu$ are the Lagrange multipliers corresponding to the amplitude equations; $\frac{\partial \mathcal{L}_{\mathrm{CC}}}{\partial \lambda_\mu} = 0$ leads to the coupled cluster working equations in Eq.~\ref{eq::cc-equations}; $\frac{\partial \mathcal{L}_{\mathrm{CC}}}{\partial t_\mu} = 0$ leads to the equations for the Lagrangian multipliers. The expectation value of the observable $O$ is then,
\begin{equation}
    \langle O \rangle = \langle \Phi_0 | (1 + \Lambda_{\mathrm{CC}}) \; \mathrm{e}^{-T_{\mathrm{CC}}} O \, \mathrm{e}^{T_{\mathrm{CC}}} | \Phi_0 \rangle
\end{equation}

The mean-field optimization of the reference in Eq.~\ref{eq:mfansatz} already allows for a non-trivial mean-field state (e.g. a localized mean-field state), and we start the coupled cluster equations from both the localized and delocalized mean-field solutions. 
We also consider a further orbital optimization to make the coupled clus1ter Lagrangian stationary, corresponding to the ansatz
\begin{align}
    |\Psi_{\mathrm{CC}}\rangle = \exp(T_\mathrm{CC})|\Phi_0(\mathbf{A}, \boldsymbol{\xi})\rangle
\end{align}
where $\mathbf{A}, \boldsymbol{\xi}$ are the parameters in Eq.~\ref{eq:mfansatz}, relaxed in the presence of the coupled cluster correlations, and the $T_\mathrm{CC}$ amplitude are also updated with the orbital parameters. We refer to this mean-field reference as $|\Phi_0^{(\mathrm{CC})}\rangle$. (In practice, we relax $\mathbf{A}$ and update $\boldsymbol{\xi}$ parametrically via $\xi_l = g|A_l|^2/\omega$ as we did in the self-consistent optimization of Eq.~\ref{eq:mfansatz}; we find in the Lang-Firsov simulations below that there is little difference between independent optimization of $\boldsymbol{\xi}$ and using this parametric choice).  


\subsection{Variational Lang-Firsov Approach and Perturbation Theory}
\label{subsec::eph-lf}
The Lang-Firsov (LF) transformation is a unitary exponential {ansatz} to obtain the 
ground state. In this study, our focus is solely on the diagonal formulation 
of the Lang-Firsov parameters \cite{cui2023},
\begin{equation}\label{eq::lf-transform-u}
|\Psi_\mathrm{LF}\rangle = \exp(T_{\mathrm{LF}} - T^\dagger_{\mathrm{LF}}) |\Phi_0\rangle 
= U_{\mathrm{LF}} |\Phi_0\rangle
\end{equation}
\begin{equation}
T_{\mathrm{LF}} = \sum_{l} \lambda_l a^\dagger_l a_l b^\dagger_l
\end{equation}
and the energy is 
\begin{align}
    E_\mathrm{LF}(\bm{\lambda}) = \langle \Phi_0  | U^{\dagger}_\mathrm{LF}(\boldsymbol{\lambda}) \, H 
    \, U_\mathrm{LF}(\boldsymbol{\lambda})
    |\Phi_0\rangle  
    = \langle\Phi_0 | {H}_\mathrm{LF}(\boldsymbol{\lambda})
    |\Phi_0 \rangle
\end{align}
where ${H}_\mathrm{LF}(\boldsymbol{\lambda})$ is the Lang-Firsov Hamiltonian. 
In the coupled cluster classification this ansatz is a unitary coupled cluster model containing only a type of first-order $T_{\mathrm{ep}}$ operator, i.e. a variant of unitary CC0-0-S1, although we refer to it as LF below.

Just as with the coupled cluster ansatz, the LF energy can be computed from different choices of $|\Phi_0\rangle$, and we consider both the delocalized and localized mean-field solutions of Eq.~\ref{eq:mfansatz}. In addition, we also reoptimize the reference state $|\Phi_0\rangle$ in the presence of the electron-phonon correlations. This corresponds to defining
\begin{equation}
\begin{aligned}
    E_{\mathrm{LF}} (\boldsymbol{\lambda}, \mathbf{A}, \boldsymbol{\xi}) 
    & = \langle\Phi_0 (\mathbf{A}, \boldsymbol{\xi}) | {H}_\mathrm{LF}(\boldsymbol{\lambda})
    |\Phi_0 (\mathbf{A}, \boldsymbol{\xi})\rangle \\    
\end{aligned} \label{eq:lfmf}
\end{equation}
and using the analytic gradients obtained in Ref.~\cite{cui2023}, we minimize $E_\mathrm{LF}$ with respect to all the parameters. In Ref.~\cite{cui2023}, this fully optimized Lang-Firsov state is referred to as the Lang-Firsov Hartree-Fock (LF-HF) energy. 

\begin{figure*}[htb]
    \centering
    \includegraphics[width=0.8\textwidth]{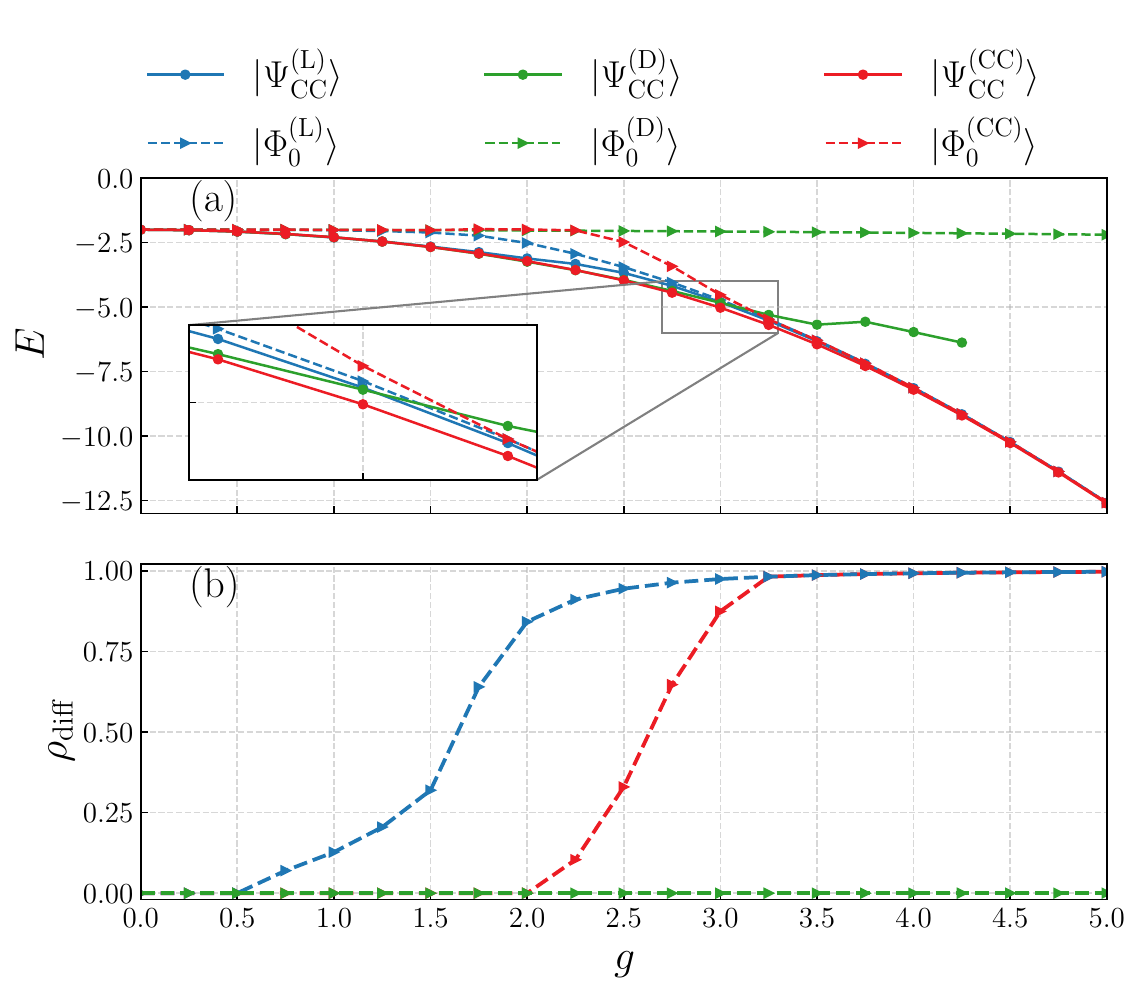}
    \caption{The ground-state energy of the 64-site Holstein model with $\omega = 2.0$ and $0 \leq g \leq 5.0$
    based on different mean-field reference states.
    The dashed lines represent the energies of the reference states, and the solid lines represent the energies of the corresponding exponential ansatz. 
    In (a), the energies of the reference states and that of CCS-2-S2 are shown, the inset shows the energies near the transition region;
    in (b), the density difference $(\rho_\text{max}-\rho_\text{min})$ of the reference states is plotted.
    }
    \label{fig:orbital-opt}
\end{figure*}


To incorporate additional electron-phonon correlation, we can carry out perturbation theory.
\newpage

Similar to the conventional M\o{}ller-Plesset (MP) theory utilized in electronic structure theories, we define the
zeroth-order Hamiltonian as
\begin{equation}
{H}^{(0)}_{\mathrm{LF}} = \bar{F} + \omega \sum_{l} \qty(\xi^2_l + b^{\dagger}_l b_l) 
\end{equation}
where $\bar{F}$ is the effective electronic Hamiltonian $\langle \Phi_0^{\mathrm{B}}|H^{\mathrm{LF}}|\Phi_0^{\mathrm{B}}\rangle$, where $|\Phi_0^{\mathrm{B}}\rangle$ is the phonon part of the state defined  in Eq.~\ref{eq:mfansatz}.
$|\Phi_0\rangle$ is an eigenstate of ${H}^{(0)}_{\mathrm{LF}}$ and the corresponding energy is
$E^{(0)}_{\mathrm{LF}} = \bar{F}_{00} + \omega \sum_{l} \xi^2_l$ .

The fluctuation potential is then
\begin{equation}
W = {H}_\mathrm{LF} - {H}^{(0)}_{\mathrm{LF}}
\end{equation}
The corresponding first-order energy correction is
\begin{equation}
E^{(1)}_{\mathrm{LF}} = \mel{\Phi_0}{W}{\Phi_0} = E_{\text{LF}} - \bar{F}_{00}
\end{equation}
\begin{equation}
E^{\mathrm{MP1}}_{\mathrm{LF}} = E^{(0)}_{\mathrm{LF}} + E^{(1)}_{\mathrm{LF}}  = E_{\text{LF-HF}} + \omega  \sum_{l} \xi^2_l
\end{equation}
which is exactly the LF energy.
We then consider the second-order energy correction,
\begin{equation}
E^{(2)}_{\mathrm{LF}} = -\sum_{\mu} \frac{\qty|\mel{\Phi_0}{W}{\Phi_\mu}|^2}{E^{(0)}_\mu - E^{(0)}_0} \label{eq:lfmp2}
\end{equation}
where $|\Phi_\mu\rangle = \tau_\mu |\Phi_0\rangle$, as described in Eq.~\ref{eq::t-and-tau}. {We evaluate it in a space of electron-phonon configurations following Ref.~\cite{cui2023}}, which requires a truncation of the phonon number. Here, we choose to truncate at 10 phonons per site.


\clearpage
\section{Results and Discussions}
\label{sec::Results}
We apply the methods described in Section~\ref{sec::Theory} to the one-dimensional Holstein model within the parameter space of $0.1 \leq \omega \leq 4.0$ and $0 \leq g \leq 5$ in units of $t$. We use converged DMRG results as the reference, and the convergence of this data 
is discussed in Appendix~\ref{appendix::eph-dmrg}.

\subsection{Role of the reference}
\label{sec:reference}
We first discuss the role of the reference state for the coupled cluster method 
in describing the physics of the Holstein model. 
We consider the following choices of reference state for the coupled cluster method:
(i) the lowest energy optimized mean-field state allowing for (potential) breaking of translational invariance. This reference state is denoted  $|\Phi_0^{(\mathrm{L})}\rangle$
and the corresponding exponential ansatz as $|\Psi_{\mathrm{CC}}^{(\mathrm{L})}\rangle$; 
(ii) the delocalized reference state with $A_l = 1 / \sqrt{L}$  and $\xi_l = g / L \omega$, denote as  $|\Phi^{(\mathrm{D})}_{0}\rangle$ with the corresponding $|\Psi_{\mathrm{CC}}^{(\mathrm{D})}\rangle$;
and (iii) the reference state that minimizes the coupled cluster energy, denoted as $|\Phi_0^{(\mathrm{CC})}\rangle$ with the corresponding $|\Psi_{\mathrm{CC}}^{(\mathrm{CC})}\rangle$.

We present results from the 64-site Holstein model with $\omega = 2.0$.
Fig.~\ref{fig:orbital-opt} (a) shows the energy of the reference states (dashed lines) and the 
corresponding CCS-2-S2 energies (solid lines); Fig.~\ref{fig:orbital-opt} (b) plots the density 
difference
$\rho_{\mathrm{diff}} = (\rho_{\max} - \rho_{\min})$ of the reference states as 
a descriptor of the localization of the reference states.
Note that for any coupling constant $g$, the exact solution predicts a uniform electron density in the ground state. 
All reference states coincide for small values of $g$. However, 
as $g$ increases beyond a critical $g$, $|\Phi_0^{(\mathrm{L})}\rangle$ and $|\Phi_0^{(\mathrm{CC})}\rangle$
start to transform into localized states with lower energies than the delocalized reference state 
$|\Phi_0^{(\mathrm{D})}\rangle$.
It is important to note that although $|\Phi_0^{(\mathrm{D})}\rangle$
can be obtained for all $g$, we are unable to converge the coupled cluster amplitude equations in the strong coupling regime and thus cannot obtain
$|\Psi_{\mathrm{CC}}^{(\mathrm{D})}\rangle$.
The presence of electron-phonon correlation in the optimization of the reference state in $|\Phi_0^{(\mathrm{CC})}\rangle$
shifts the critical $g$ towards a larger value.  The density differences in Fig.~\ref{fig:orbital-opt} (b) also reflect a smaller amount of symmetry breaking in the $|\Phi_0^{(\mathrm{CC})}\rangle$ compared to $|\Phi_0^{(\mathrm{L})}\rangle$.
Interestingly, near the transition ($1.0 \leq g \leq 3.5$) the lowest energy mean-field reference 
does not result in the lowest energy coupled cluster energy.
As shown in the zoomed-in plot, while $|\Phi_0^{(\mathrm{L})}\rangle$ corresponds to the lowest mean-field energy,
the energies of both $|\Psi_{\mathrm{CC}}^{(\mathrm{D})}\rangle$ and $|\Psi_{\mathrm{CC}}^{(\mathrm{CC})}\rangle$
are lower.

The above illustrates the importance of choosing an appropriate reference when using an exponential ansatz. In the sections below, we always use the optimized reference state (iii) for the coupled cluster methods, and perform the full reoptimization of $|\Phi_0\rangle$ in the LF-HF and LF-MP2 calculations.

\begin{figure}[htbp]
    \centering
    \includegraphics[width=0.48\textwidth]{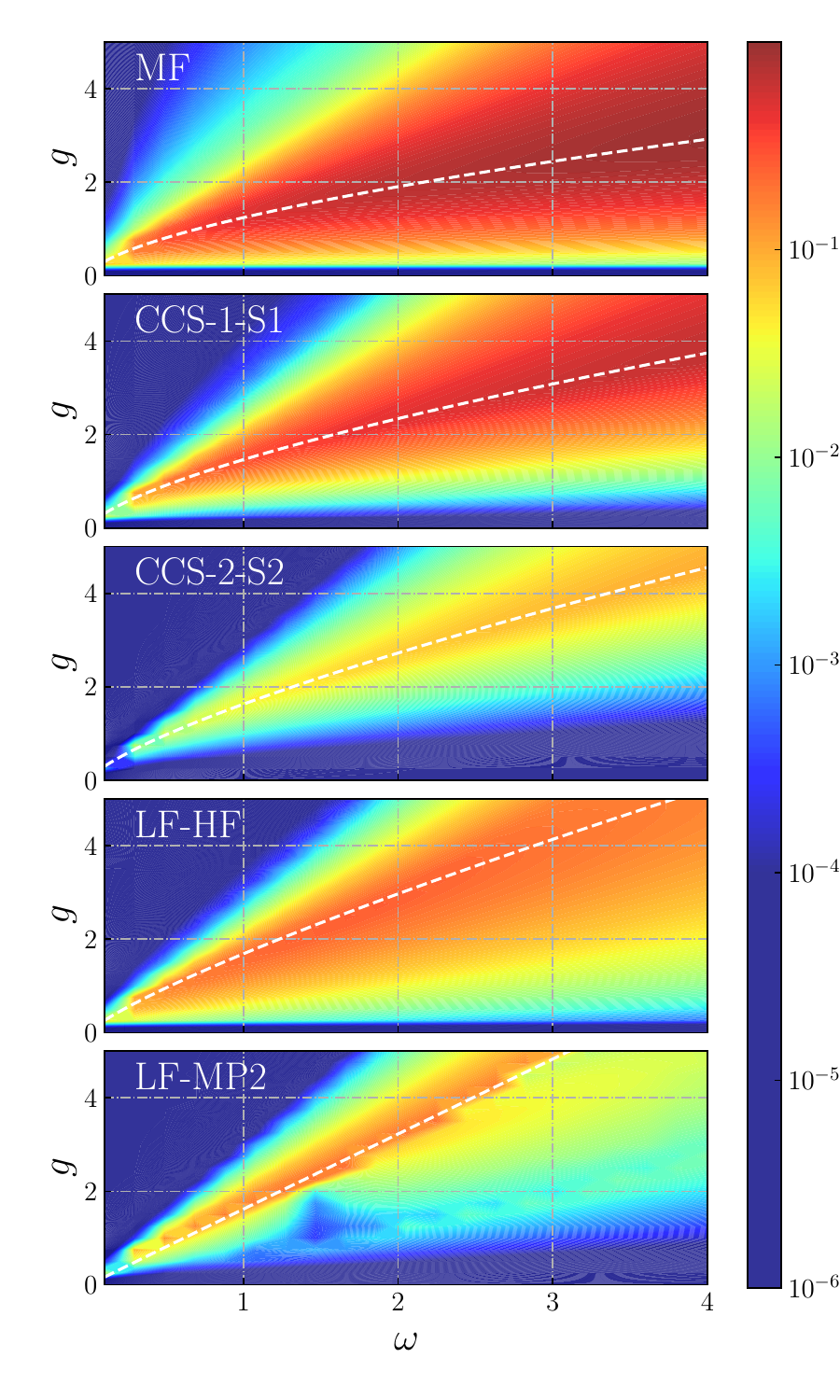}
    \caption{The error of the ground state energy (units of $t$) in the one-dimensional 64-site Holstein model using different methods for $0.1 \leq \omega \leq 4.0$ and $0 \leq g \leq 5$. The white dashed line shows the coupling strength at the given frequency 
    where the method exhibits its largest error.
    MF is defined in Eq.~\ref{eq:mfansatz}, the exponential ansatzes are defined in Table~\ref{tab::models}, and LF-MP2 is defined in Sec.~\ref{subsec::eph-lf}.
    }
    \label{fig::hol-64-err-2d}
\end{figure}

\subsection{Energy across parameter space}
\begin{figure*}[htbp]
    \centering
    \includegraphics[width=0.90\textwidth]{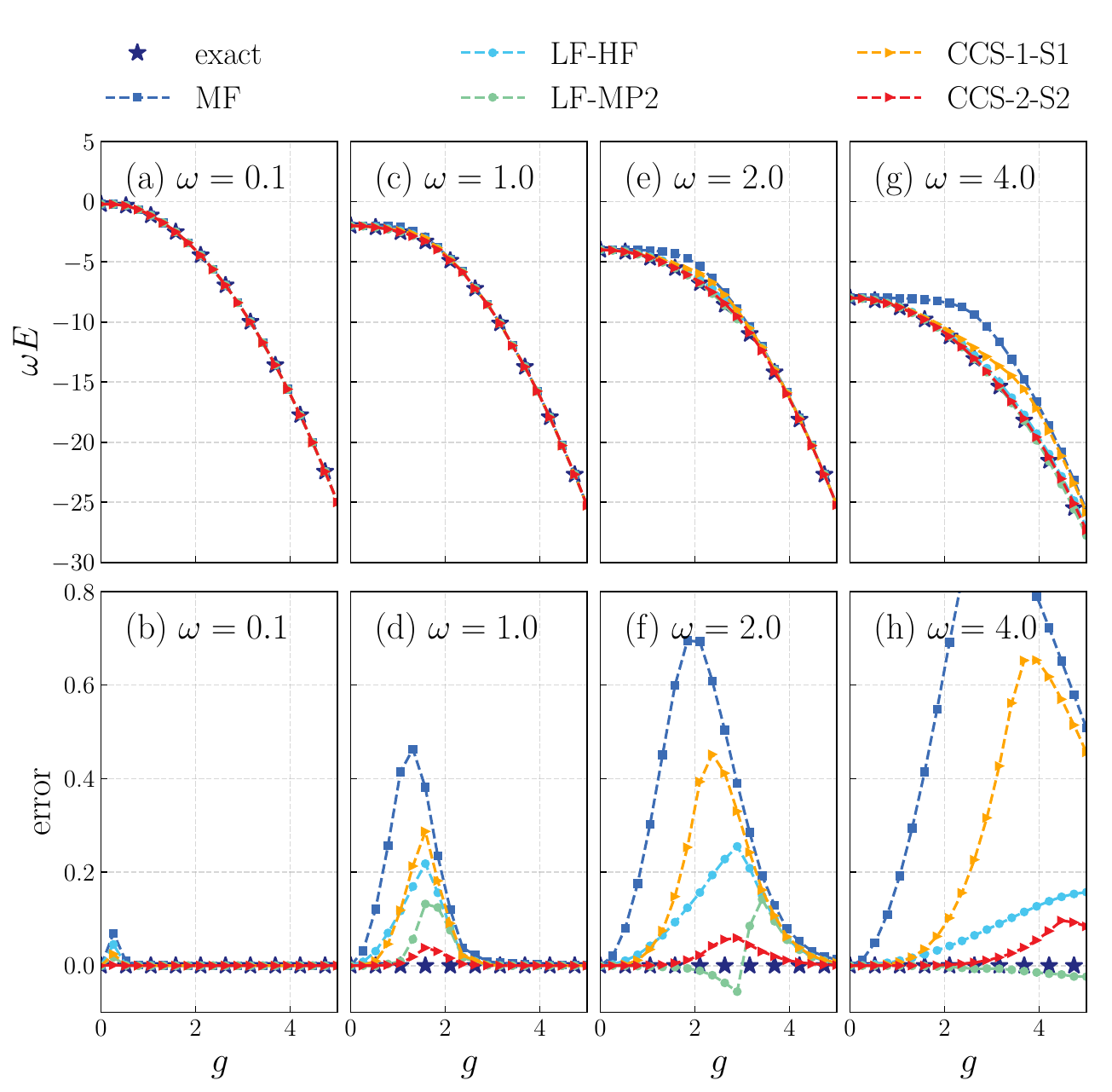}
    \caption{
        $\omega$ slices in parameter space for the scaled ground state energy (a, c, e, and g) and the error (b, d, f, and h) in the one-dimensional 64-site Holstein model. Method labels are the same as in Fig.~\ref{fig::hol-64-err-2d}.
        (a) and (b) are for $\omega = 0.1$; (c) and (d) are for $\omega = 0.5$;
        (e) and (f) are for $\omega = 1.0$; (g) and (h) are for $\omega = 2.0$.
    }
    \label{fig::hol-64-err-slice}
\end{figure*}

We now examine the energy errors across the Holstein parameters space for Lang-Firsov and coupled cluster ansatzes. 
Fig.~\ref{fig::hol-64-err-2d} plots the ground state energy error in the 2D parameter space of $\omega$ and $g$ for the 64-site Holstein model, while slices through this space are shown in Fig.~\ref{fig::hol-64-err-slice} for $\omega = 0.1, 0.5, 1.0$ and $2.0$. All methods presented are accurate for $g=0$ and in the strong coupling regimes, and display most variation in accuracy in the intermediate coupling regime. 
The dashed lines in Fig.~\ref{fig::hol-64-err-slice} represent the coupling strength at the given frequency where the method exhibits its largest error.
From this we can conclude that the accuracy roughly follows MF$<$CCS-1-S1$\approx$LF-HF$<$LF-MP2$<$CCS-2-S2. Similar conclusions can be drawn from Fig.~\ref{fig::hol-64-err-slice}. These slices additionally show that the mean-field error is largest in the anti-adiabatic ($\omega=2.0$) regime. Despite the similarity of the LF-HF and CCS-1-S1 parametrizations, LF-HF outperforms CCS-1-S1 in this regime: the unitary variational optimization of LF-HF clearly contributes to the improved behaviour in this limit. We see also that LF-MP2 has a discontinuity in the energy at intermediate decoupling, which does not appear in the coupled cluster results.

\subsection{Electronic Kinetic Energy}
\begin{figure*}[htbp]
    \centering
    \includegraphics[width=0.9\textwidth]{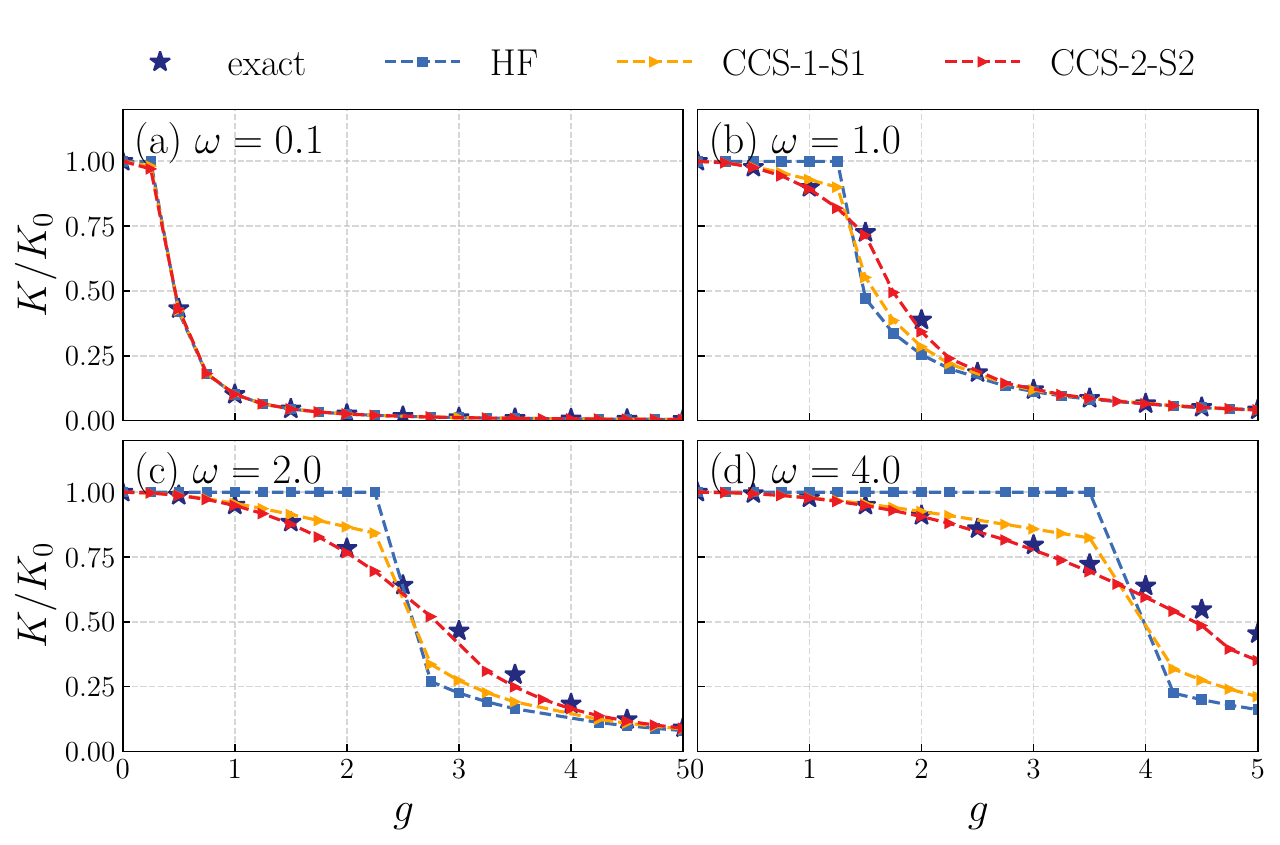}
    \caption{The kinetic energy (in units of $K_0$) of the electron in the one-dimensional 16-site Holstein model. The star symbols represent the exact solution and the dashed lines with 
    symbols represent the approximate numerical methods. Method labels are the same as in Fig.~\ref{fig::hol-64-err-2d}.
    (a) $\omega = 0.1$; (b) $\omega = 0.5$; (c) $\omega = 1.0$; and (d) $\omega = 2.0$.
    }
    \label{fig::hol-16-k}
\end{figure*}

The ratio of the electron kinetic energy ($K$) to its kinetic energy at zero electron-phonon coupling ($K_0$) provides a simple way to diagnose the self-trapping transition even in an exact calculation where the ground-state density remains uniform.
Fig.~\ref{fig::hol-16-k} plots $K / K_0$ as a function of the electron-phonon coupling strength $g$ in a 16-site Holstein model for the coupled cluster methods.
All the methods accurately capture both the strong and weak coupling regimes. However, as previously noted, the simple mean-field method results in a discontinuity. 
As additional electron-phonon coupling is included, in CCS-1-S1 and CCS-2-S2, the 
continuity of the transition is gradually restored. Indeed, the CCS-2-S2 result is visually indistinguishable from the exact reference.


\subsection{Electron-Lattice Correlation}
\begin{figure*}[htbp]
    \centering
    \includegraphics[width=1.0\textwidth]{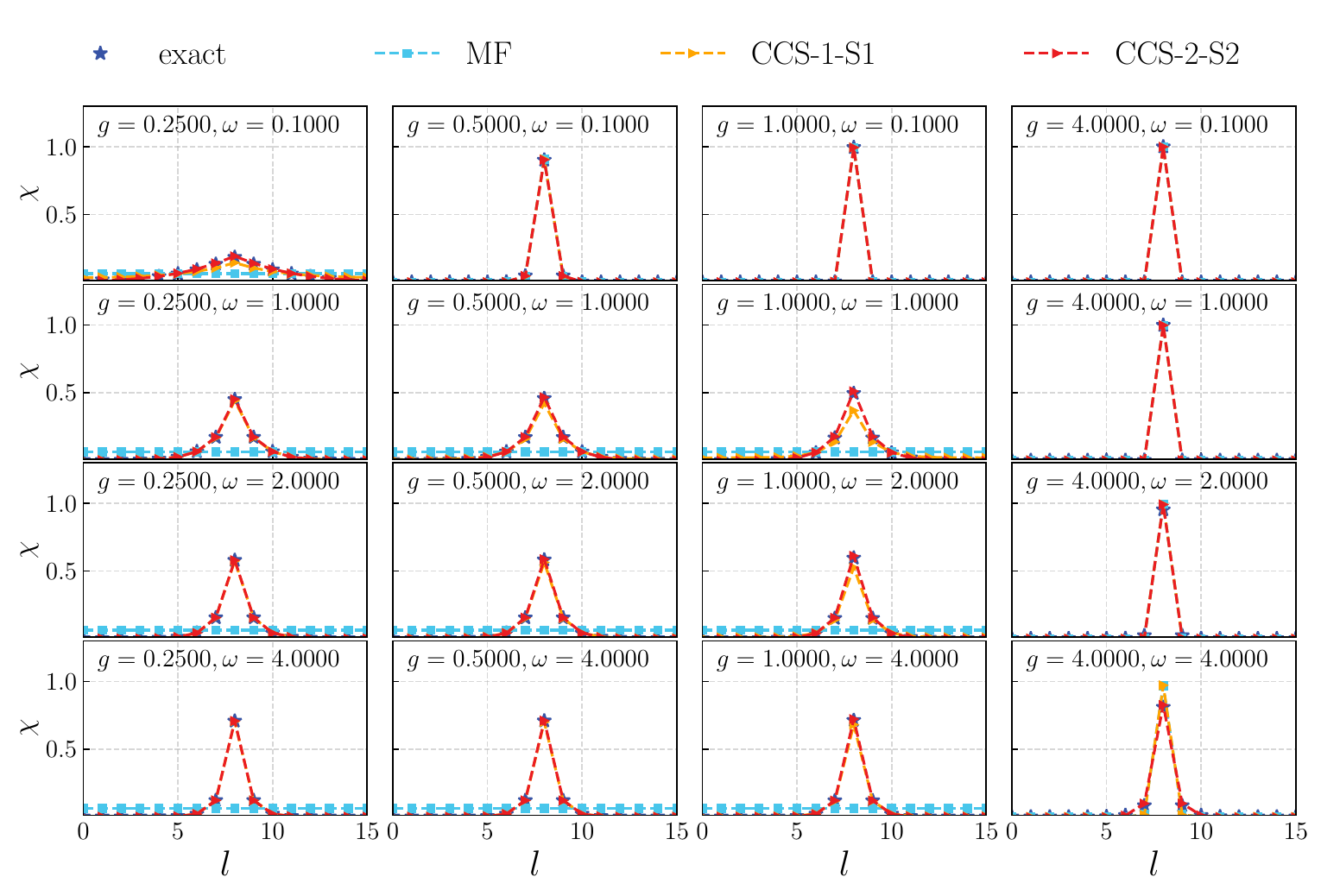}
    \caption{Electron-lattice correlation function $\chi$ at site 8 in the 16-site Holstein model for various 
    values of $\omega$ and $g$ annotated in the figure. Method labels as in Fig.~\ref{fig::hol-64-err-2d}.}
    %
    \label{fig::hol-16-chi}
\end{figure*}


 

Fig.~\ref{fig::hol-16-chi} displays the 16-site Holstein model's normalized electron-lattice correlation function for various $\omega$ and $g$ values for the MF and CC approximations. (We choose $k=8$, the middle of the lattice). Note that although the MF solution localizes the electron, there is no localization in the electron-phonon correlation function as $|\Phi_0\rangle$ is a product state, i.e. $\chi_{lk}=\rho_k$. As the inclusion of electron-phonon correlation increases from MF to CCS-1-S1 and CCS-2-S2, the electron-phonon correlation function becomes more compact, even as the spatial localization of the optimized reference state decreases (see Sec.~\ref{sec:reference}).

\section{Conclusions}
\label{sec::Conclusions}
We have benchmarked the exponential ansatz across the parameter space of the Holstein model, comparing to near-exact results from DMRG.
Allowing for a relaxation of the reference state that is the starting point for the exponential ansatz, we find that we can obtain a good description in the weak and strong coupling regimes, corresponding to large and small polarons. Within the systematic coupled cluster framework, increasing the electron-phonon excitation level leads to increasingly accurate results both for the energy and the correlation functions, and at the doubles level of approximation, these results are often visually indistinguishable from the exact ones. The variational Lang-Firsov transformation performs better than the lowest (i.e. singles) rung of the coupled cluster approximation hierarchy, particularly in the intermediate coupling regime. 

Looking beyond models, the exponential ansatz is the starting point for a large number of applications in accurate ab initio electronic structure. The results here suggest that it is a competitive approach for polaron physics, providing a viable path forward to describe correlated electron-phonon effects at the ab initio many-body level. Another interesting direction is to examine the application of exponential wavefunctions such as the coupled cluster hierarchy to systems of many-electrons and phonons, along the lines first discussed in Ref.~\cite{gao2020}. The incorporation of more flexible references, such as superconducting states, is of particular interest there.

\begin{acknowledgments}
This work was supported by the U.S. Department of Energy, Office of Science,
Office of Advanced Scientific Computing Research and Office of Basic Energy Sciences, Scientific Discovery through Advanced Computing (SciDAC) program under Award Number DE-SC0022088. We thank Shuoxue Li and Yao Luo for helpful discussion; and
Dr. Alec White for providing valuable assistance in utilizing the \textsc{Wick} package.
Data and codes used in this work can be obtained in \url{https://github.com/yangjunjie0320/exp-ansatz-holstein-model}.
\end{acknowledgments}

\appendix
\section{Density Matrix Renormalization Group Theory}
\label{appendix::eph-dmrg}
\begin{figure*}[htbp]
    \centering
    \includegraphics[width=1.0\textwidth]{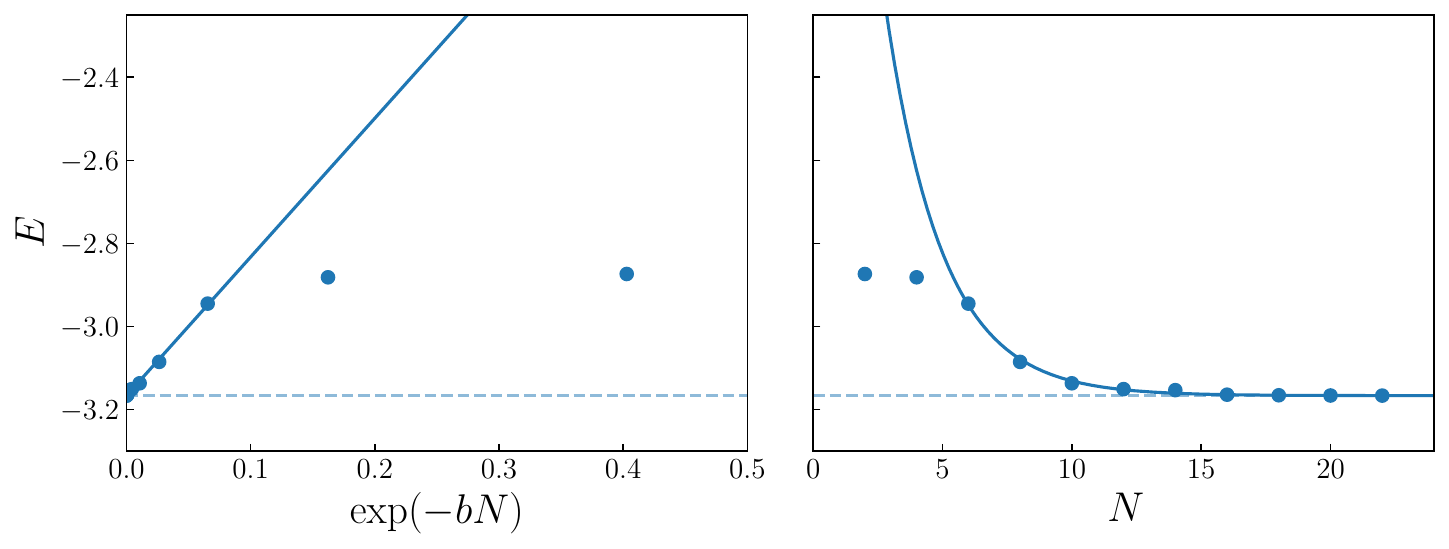}
    \caption{Extrapolation of the ground state energy of the 64-site Holstein model with respect to the phonon truncation for $\omega = 0.5$ and $g = 1.15$. The extrapolation is 
    performed with the function $E(N) = E_{\infty} + a \exp(-b N)$.}
    \label{fig::hol-64-dmrg-extrapolate}
\end{figure*}
The density matrix renormalization group (DMRG) is a numerically exact method for solving the quantum many-body problem.
The method is particularly accurate and efficient for one-dimensional systems~\cite{white1989,white1993}. 
In early developments, DMRG was applied to the Holstein model in Ref.~\cite{jeckelmann1998} to calculate the ground state energy and the electron-lattice correlation function, and to calculate the dynamical properties of the Holstein model~\cite{zhang1999a}.
From a modern perspective, the DMRG method is based on the matrix product state (MPS) representation of the coefficients
in Eq.~\ref{eq::fci},
\begin{equation}
    C_I = \mathbf{A}^{\sigma_1^I} \cdot \mathbf{B}^{n_1^I} \cdot \mathbf{A}^{\sigma_2^I} \cdot \mathbf{B}^{n_2^I} \cdot \cdots \cdot \mathbf{A}^{\sigma_L^I} \cdot \mathbf{B}^{n_L^I}
\end{equation}
in which $\mathbf{A}^{\sigma_l^I}$ and $\mathbf{B}^{n_l^I}$ are matrices with a maximum bond dimension $D$, representing the electronic ($\sigma$) and phonon ($n$) degrees of freedom in the site basis, respectively. 
The Hamiltonian is first transformed to the shifted-phonon basis of the lowest mean-field state of Eq.~\ref{eq:mfansatz}, corresponding to 
\begin{align}
    b_l^{\dag} &\to b_l^{\dag} + \xi_l^* \quad b_l \to b_l + \xi_l \nonumber\\
    H &\to  -t \sum_{\langle kl \rangle}  a_{k}^{\dagger} a_{l} - 2 g \sum_l \xi_l a_{l}^{\dagger} a_{l}
     + g \sum_l \rho_l\left(b_l^{\dagger} + b_l\right)   \notag\\ 
    & +  \omega \sum_l b_l^{\dagger} b_l - \omega \sum_l \xi_l \left(b_l^{\dagger} + b_l\right) 
\end{align}
and this Hamiltonian is expressed as a matrix product operator (MPO) in the site basis, 
$H = \sum_{IJ} H_{IJ} |\Phi_I\rangle \langle \Phi_J|$ with, 
\begin{equation}
    H_{IJ} = \mathbf{W}^{\sigma_1^I\sigma_1^J} \cdot \mathbf{U}^{n_1^In_1^J} \cdot \cdots \cdot \mathbf{W}^{\sigma_L^I\sigma_L^J} \cdot \mathbf{U}^{n_L^In_L^J}
\end{equation}
We then optimize the energy expectation value,
\begin{equation}
    E = \frac{\langle \Psi_{\mathrm{MPS}} | H_{\mathrm{MPO}} | \Psi_{\mathrm{MPS}} \rangle}{\langle \Psi_{\mathrm{MPS}} | \Psi_{\mathrm{MPS}} \rangle}
\end{equation}
and converge our calculations with respect to the bond dimension $D$ and (shifted) phonon cutoff $N$. Fig.~\ref{fig::hol-64-dmrg-extrapolate} shows the ground state energy of the 64-site Holstein model with respect to the phonon cutoff for $\omega = 0.5$ and $g = 1.15$. The energy is extrapolated to the infinite phonon cutoff limit using the function $E(N) = E_{\infty} + a \exp(-b N)$, where $E_{\infty}$ is the extrapolated energy. The bond dimension was kept at $D=500$ for both electron and boson sites.
The electron-phonon DMRG implementation is available through the
\textsc{Block2} package~\cite{zhai2021,zhai2023}.

\vspace{5mm}
\bibliography{bibliography.bib}
\end{document}